\documentclass[twocolumn,floatfix,prl,aps,showpacs]{revtex4}

\usepackage{graphicx}
\usepackage{bm}
\usepackage{amsmath}
\usepackage{amsfonts}

\newcommand{\Dslash}{{\slash{\kern -0.5em}\partial}}
\newcommand{\Aslash}{{\slash{\kern -0.5em}A}}

\def\sqr#1#2{{\vcenter{\hrule height.#2pt
     \hbox{\vrule width.#2pt height#1pt \kern#1pt
        \vrule width.#2pt}
     \hrule height.#2pt}}}

\def\thinspace{\kern .16667em}

\def\xp{x_{{\kern -.2em}_\perp}}
\def\subp{_{{\kern -.2em}_\perp}}

\def\be{\begin{equation}}
\def\ee{\end{equation}}

\begin{document}

\title{Fractional quantum Hall effect in graphene}
\author{Csaba~T\H oke, Paul~E.~Lammert, Vincent~H.~Crespi, and Jainendra~K.~Jain}
\affiliation{Department of Physics, 104 Davey Lab, 
Pennsylvania State University, University Park PA, 16802}
\date{\today}

\begin{abstract} 
  Unlike regular electron spin, the pseudospin degeneracy of Fermi
  points in graphene does not couple directly to magnetic field.
  Therefore, graphene provides a natural vehicle to observe the
  integral and fractional quantum Hall physics in an elusive limit
  analogous to zero Zeeman splitting in GaAs systems.  This limit can
  exhibit new integral plateaus arising from interactions, large
  pseudoskyrmions, fractional sequences, even/odd numerator effects,
  composite-fermion pseudoskyrmions, and a pseudospin-singlet
  composite-fermion Fermi sea.  The Dirac nature of the $B=0$
  spectrum, which induces qualitative changes in the overall spectrum,
  has no bearing on the fractional quantum Hall effect in the $n=0$
  Landau level of graphene.  The second Landau level of graphene is
  predicted to show more robust fractional quantum Hall effect than
  the second Landau level of GaAs.
\end{abstract}

\maketitle

The discovery of integral quantum Hall plateaus in graphene
\cite{{Novoselov}} poses several questions: Does the fractional
quantum Hall effect (FQHE) also occur in graphene, and, if so, what is
its character?  Does it resemble the integral quantum Hall effect
(IQHE) in graphene?  How does it compare to the FQHE in GaAs?  What
other physics does interaction produce?  Even if the Zeeman energy is
large enough to fully polarize the (real) spin of the low energy
state, as assumed throughout this article, the graphene Fermi point
degeneracy, which takes the form of a pseudospin $1/2$ character,
remains.  Strikingly, despite the qualitative changes in the Landau
level (LL) spectrum that arise from the Dirac dispersion of the low
energy states at zero field, the physics of the $n=0$ graphene Landau
level is identical to that of electrons in the $n=0$ GaAs Landau
level, but with the Zeeman splitting set identically to zero; the
pseudospin, which is also a sublattice index for $n=0$, plays the role
of the traditional electron spin.  This leads to the prediction of the
formation of composite fermions, FQHE at $\nu=n/(2pn\pm 1)$, odd
versus even numerator effects, and a pseudospin singlet composite
fermion Fermi sea.  Giant pseudoskyrmions are predicted for the
ferromagnetic integral and fractional quantum Hall states.  Our
numerical diagonalizations show that, interestingly, the FQHE of
graphene in the $n=1$ Landau level is more robust than that in the
$n=1$ GaAs Landau level, due to more favorable interaction
pseudopotentials.  We assume below that the pseudospin symmetry is
exact, and also that LL mixing is negligible.

In graphene, the two Fermi points, each with a two-fold band
degeneracy, can be described by a low-energy continuum approximation
with a four-component envelope wave function whose components are
labelled by a Fermi-point pseudospin $= \pm 1$ and a
sublattice.  
The Hamiltonian for one pseudospin component is
\cite{shon-ando,Semenoff}
\begin{equation}
H=v_F
\begin{pmatrix}
0 & \Pi_x-i\Pi_y \\
\Pi_x+i\Pi_y & 0 \\
\end{pmatrix},
\end{equation}
where $v_F$ is the Fermi velocity and $\vec\Pi=\vec p+(e/c)\vec
A$.  Letting $z=x-iy$, choosing the symmetric gauge $\vec
A=\left(-{By}/{2},{Bx}/{2}\right)$, and introducing the Landau
level raising and lowering operators, $a^\dag=
\left({\bar z}/{2} - 2\partial_z\right)/\sqrt{2}$ and
$a=\left({z}/{2} + 2\partial_{\bar z}\right)/\sqrt{2}$,
the Hamiltonian becomes
\begin{equation}
H=\frac{\sqrt{2}\hbar v_F}{il_B}
\begin{pmatrix}
0 & a\\
-a^\dag & 0
\end{pmatrix}.
\end{equation}
The eigenvalue problem is conveniently formulated in terms of $H^2$, which 
can be represented as LL number operators:
\begin{equation}
H^2=\frac{2\hbar^2 v_F^2}{l_B^2}
\begin{pmatrix}
a^\dag a+1 & 0\\
0 & a^\dag a
\end{pmatrix}\;.
\end{equation}
Any $\Psi=\begin{pmatrix}\alpha\eta_{n-1,m_1} \\
  \beta\eta_{n,m_2}\end{pmatrix}$ is an eigenvector of $H^2$,
where $\eta_{n,m}$ are the standard
LL eigenfunctions in GaAs ($n=0,1,\cdots$ is the LL index,
and $m$ is the angular momentum index).
$H\Psi=E_n\Psi$ requires $m_1=m_2$ and fixes $\beta/\alpha$,
with the result that
normalized eigenvectors take the form ($\textrm{sgn}(0) = 0$ by convention)
\begin{equation}
\Psi^{(n\neq0,m)}=\frac{1}{\sqrt 2}
\begin{pmatrix}
-{\rm sgn}(n)i\eta_{|n|-1,m} \\
\eta_{|n|,m}
\end{pmatrix},\;
\end{equation}
\begin{equation}
\Psi^{(0,m)}=
\begin{pmatrix}
0 \\ \eta_{0,m}
\end{pmatrix}\;,
\end{equation}
\begin{equation}
E_n={\rm sgn}(n)\sqrt{{2\hbar v_F^2 eB|n|}/{c}}.
\end{equation}

Before discussing the FQHE\cite{Tsui82}, we describe
how  interactions are expected to affect the integral QHE in graphene.
For noninteracting electrons, when both the spin and the pseudospin
degeneracies are present, the Hall plateaus have been predicted and seen at
\be
R_H=h/(je^2)\;, \;\; j=\ldots, -6, -2, 2, 6, 10, 14, \ldots 
\ee
For magnetic fields large enough to resolve the spin bands
the Hall plateaus occur at 
\be
R_H=h/(je^2)\;, \;\; j=\ldots, -2, -1, 0, 1, 2, 4, 6, 8, \ldots
\ee 
\newline where $j=0$ implies a band insulator.  However, {\it all} integral
plateaus should, in principle, become possible for interacting electrons.
Consider a magnetic field large enough to lift the spin degeneracy at
a filling $\nu=2n+1$ with $n=1,2,3,\ldots$ where no QHE occurs for
noninteracting electrons.  For interacting electrons, the rotational
symmetry in pseudospin space is spontaneously broken due to pseudospin
exchange, with the pseudospin magnetization picking an arbitrary
direction.  While this system supports gapless pseudospin-wave
excitation, the {\it charged} excitations have a gap, thus producing a
Hall plateau.

\noindent {\bf FQHE in the $n=0$ Landau level of graphene:}
The unusual electronic dispersion of graphene around $E=0$ is
reflected in both the structure of Landau levels and in the offset of
the Hall conductance staircase.  Nonetheless, the actual wave functions of
electrons in the $n=0$ manifold are identical to those in the
conventional lowest Landau level of GaAs.  Consequently, the FQHE
in the $n=0$ Landau level of graphene at a high magnetic field maps onto
FQHE of electrons in GaAs with zero Zeeman energy\cite{Wu93}, as
corroborated and extended by the numerical results described below.

The basic physics in the $n=0$ graphene LL, therefore, is the same as
in GaAs.  Each electron captures an even number ($2p$) of quantized
vortices to become a composite fermion\cite{Jain89}.  The Berry phases
generated by the vortices effectively cancel part of the external
magnetic field, so that the dynamics of composite fermions are
governed by a reduced magnetic field $B^*=B-2p\rho\phi_0$, where
$\rho$ is the particle density in the $n=0$ Landau level and
$\phi_0=hc/e$ is the flux quantum.  The $n=0$ Landau level of
electrons splits into Landau-like levels of composite fermions, whose
filling factor $\nu^*$, in terms of the electron filling $\nu$, is
given by $\nu=\nu^*/(2p\nu^*\pm 1)$.

\begin{table}[b]
\begin{center}
\begin{tabular}{|r|r|r|r|r|c|c|c|c|c|c|c|}
\hline\hline
$\nu$ & $N$ & $2Q$ & $D$ & $D_0$ & $L^{(0)}$ & $S^{(0)}$ & 
\multicolumn{2}{c|}{GaAs} & \multicolumn{2}{c|}{graphene} & overlap\\ 
\cline{8-11}
& & & & & & & $L^{(1)}$ & $S^{(1)}$ & $L^{(1)}$ & $S^{(1)}$ & \\ 
\hline
$1/3$
& 4  & 9  &    145 &  5 & 0 & 2 & 0 & 2 & 0 & 2 & 0.99932\\
& 5  & 12 &   1106 & 10 & 0 &${5}/{2}$ & 2 &${3}/{2}$ & 0 &${5}/{2}$ & 0.99998\\
& 6  & 15 &  11588 & 50 & 0 & 3 & 0 & 3 & 0 & 3 & 0.99875\\
& 7  & 18 & 109138 &290 & 0 &${7}/{2}$ & 0 &${7}/{2}$ & 0 &${7}/{2}$ & 0.99974\\
\hline
$2/3$
& 4  &  5 &     29 &  3 & 0 & 0 & 0 &2 & 0 & 0 & 0.99968\\
& 6  &  8 &    500 & 10 & 0 & 0 & 1 &3 & 0 & 0 & 0.98887\\
& 8  & 11 &  11483 & 91 & 0 & 0 & 2 &4 & 0 & 0 & 0.97482\\
\hline
$2/5$
& 4  &  7 &     72 &  4 & 0 & 0 & 0 & 0 & 0 & 0 & 0.99248\\
& 6  & 12 &   3796 & 28 & 0 & 0 & 1 &3 & 0 & 0 & 0.93622\\
& 8  & 17 & 274842 &768 & 0 & 0 & 0 & 4 & 0 & 0 & 0.95578 \\
\hline
$3/5$
& 5  &  8 &    226 &  5 & 0 & ${3}/{2}$& 0 &${5}/{2}$& 0 & ${3}/{2}$& 1.00000\\
& 8  & 13 &  39131 &205 & 0 & 2 & 0 & 4 & 0 & 0 & 0\\
\hline\hline
\end{tabular}
\end{center}
\caption{\label{singlet} 
Orbital angular momentum $L^{(n)}$ and spin/pseudospin $S^{(n)}$ of the 
ground states of finite systems on a sphere at 
$\nu=\frac{1}{3},\frac{2}{3},\frac{2}{5}$ and $\frac{3}{5}$ in the $n=0$ and
$n=1$ Landau levels of graphene and GaAs.
$D$ is the dimension of the Hilbert space in the $L_z=S_z=0$ sector;
$D_0$ is the dimension of the $L=0$ sector.
The last column gives the overlaps between the $n=0$ and 
$n=1$ graphene ground states.}
\end{table}
 
The IQHE of composite fermions for $\nu^*=n$ produces sequences of
fractions\cite{notation}:
\begin{equation}
\nu=n/(2pn\pm 1)\;.
\label{eq:f}
\end{equation}
The origin of gap, i.e. the energy required to promote a composite
fermion into a higher CF-Landau level, is different for even and odd
values of $n$.  For even $n$, the ground state is a pseudospin
singlet, with $n/2$ CF-Landau levels for each component of the
pseudospin occupied.  For odd $n$, the ground state is partially
pseudospin polarized; no QHE would occur here if the composite
fermions did not interact, but the residual interaction between
composite fermions opens a gap.  To the extent the residual
interaction is weak, one expects fractions with even numerators to be
more robust than those with odd numerators.  The excitation energies
for the GaAs FQHE in the zero Zeeman energy limit apply to graphene
FQHE within the $n=0$ level.  Calculated gaps to creation of a
far-separated charged quasiparticle/quasihole pair at $\nu = 1/3$ and
(unpolarized) $\nu = 2/5$ are 0.07 and 0.04 $e^2/\epsilon \ell_B$,
respectively \cite{Rezayi2}; the larger gap at 1/3 indicates the
significance of inter-CF interactions.

So long as electrons are confined to the $n=0$ Landau level, they have
no memory of the Dirac nature of the zero-field dispersion, with some
surprising consequences for the FQHE.  The CF-cyclotron energy opens
up approximately linearly with $B^*$, as expected for composite
fermions with a parabolic dispersion, even though the cyclotron energy
of {\it electrons} in graphene scales anomalously with $B$.

In GaAs, the sequence of FQHE states at $\nu=n/(2pn\pm 1)$ terminates
as $n\rightarrow \infty$ in a composite-fermion Fermi sea at
$\nu=1/2p$, where the effective magnetic field vanishes
\cite{Kalmeyer92,Willett93}.  For zero Zeeman energy, variational
calculations favor the spin singlet Fermi sea\cite{Park}, so graphene
should have a pseudospin-singlet CF Fermi sea at $\nu=1/2p$.  The CF
Fermi sea in GaAs has been successfully modeled as an ordinary Fermi
sea with parabolic dispersion, which allows one to deduce an effective
mass for composite fermions.  The same should be true of the CF Fermi
sea in graphene, in spite of the fact that electrons in graphene make
a Dirac sea at zero magnetic field and have no effective mass.  The
singlet nature of the CF Fermi sea can be ascertained through a
measurement of the Fermi wave vector, as was accomplished in GaAs
systems by various geometric means \cite{Willett93}.

\begin{figure*}[htbp]
\begin{center}
\includegraphics[scale=0.95]{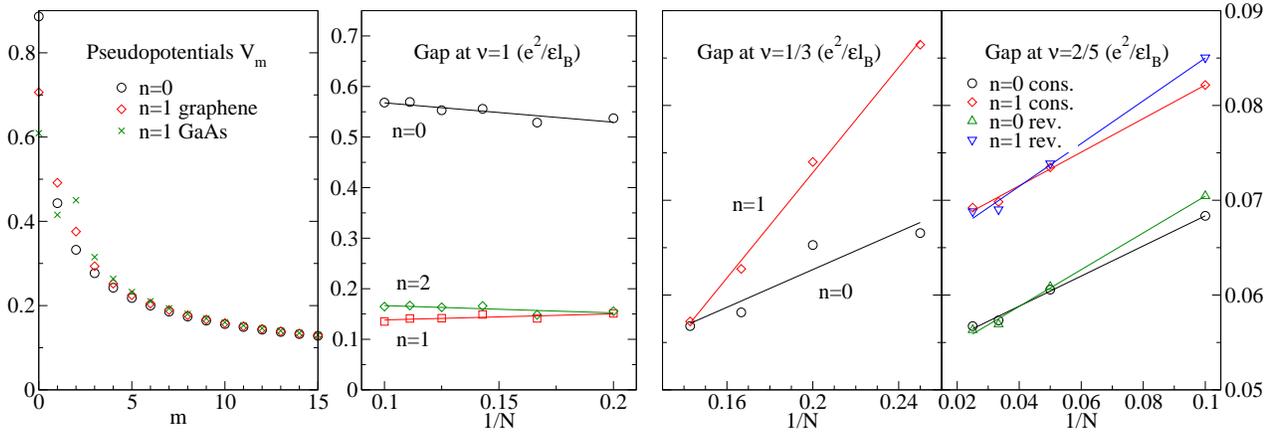}
\end{center}
\caption{\label{allgraphs} (a) Pseudopotentials for $n=0$ and $n=1$
  Landau levels in graphene and GaAs.  (b)-(d) Activation gaps at
  $\nu^{(n)}=1,1/3$ and 2/5 in several graphene LLs.  All energies are
  given in units of $e^2/\epsilon l_B$, where $\epsilon$ is the
  dielectric constant of the host semiconductor and $l_B$ is the
  magnetic length.  The gaps in (b) and (c) refer to the energy
  required to create a pseudoskyrmion-antiskyrmion pair.  Gaps to
  pseudospin conserving (cons.) and pseudospin reversed (rev.)
  excitations are given in (d).  }
\end{figure*}

\noindent {\bf FQHE in the $n=1$ Landau level of graphene:}
The nature of FQHE depends on the Haldane pseudopotentials.  In GaAs,
the FQHE is essentially restricted to the lowest LL: very few
fractions are seen in $n=1$, and almost none in higher LLs.  The
mapping between GaAs and graphene does not hold in higher Landau
levels, so different behaviors are expected. We first evaluate the 
Coulomb matrix elements within the $n$th graphene Landau level.  
Write $|n_1,m_1;n_2,m_2;\dots;n_N,m_N\rangle$ 
for the product state
$\eta_{n_1,m_1}\otimes\eta_{n_2,m_2}\otimes\dots\otimes\eta_{n_N,m_N}$
and
$\parallel n_1,m_1;n_2,m_2;\dots;n_N,m_N\rangle{\kern -2 pt}\rangle$
for $\Psi^{(n_1,m_1)}\otimes\dots\otimes\Psi^{(n_N,m_N)}$.
Then,
\begin{eqnarray}
\label{matrixelement}
&&4 \langle{\kern -2 pt}\langle n,m_1;n,m_2\parallel V\parallel n,m_3;n,m_4
\rangle{\kern -2 pt}\rangle = \notag\\
&&=\langle n,m_1;n,m_2|V|n,m_3;n,m_4\rangle+ \\
&&\langle n-1,m_1;n,m_2|V|n-1,m_3;n,m_4\rangle+ \notag\\
&&\langle n,m_1;n-1,m_2|V|n,m_3;n-1,m_4\rangle+ \notag\\
&&\langle n-1,m_1;n-1,m_2|V|n-1,m_3;n-1,m_4\rangle. \notag
\end{eqnarray}
By conservation of angular momentum, these matrix elements are all
proportional to $\delta_{m1+m2,m3+m4}$.  The problem of electrons in
the $n^{\rm th}$ graphene LL thus formally maps into that of the
lowest GaAs LL (with two pseudospin copies) with an effective
interaction defined by the pseudopotentials
\begin{equation} 
\label{effective}
V^{(n)\textrm{graphene}}_m=\frac{1}{4}
\left[V^{(n)}_m+V^{(n-1)}_m+2V_m^{(n,n-1)}\right]\;,
\end{equation} 
\begin{equation} 
V_m^{(n,n-1)}=\int\frac{d^2k}{(2\pi)^2}\frac{2\pi}{k}L_n\left(\frac{k^2}{2}\right)L_{n-1}
\left(\frac{k^2}{2}\right)e^{-k^2}L_m(k^2),
\nonumber
\end{equation}
and $V_m^{(n)}$ is the effective pseudopotential for the $n^{\rm th}$ LL in GaAs.
As seen in Fig.~\ref{allgraphs}(a), $V^{(1)\textrm{graphene}}_m$ lies 
between $V_m^{(0)}$ and $V_m^{(1)}$, except for $m=1$.  To see what
FQHE this interaction implies, we have numerically diagonalized finite
systems in a geometry where $N$ electrons move on the surface of a
sphere whose center holds a magnetic monopole of strength $Q$
producing a flux of $2Qhc/e$ through the surface of the sphere.  We
use the pseudopotentials calculated above for the disk geometry; this
gives the exact result for very large systems and is generally a
reasonable approximation.  

As a preliminary step, we have investigated the fully
pseudospin-polarized sector, where larger systems (with up to ten
particles) can be studied, and found that the $n=1$ graphene LL
behaves very similarly to the $n=0$ LL.  This suggests that the FQHE
in higher graphene LLs may be more robust than in GaAs.  We next
compare ground states for graphene systems including the pseudospin
degree of freedom (but with the real spin frozen), to GaAs systems
with zero Zeeman splitting.  Table \ref{singlet} shows the ground
state quantum numbers, orbital angular momentum $L$ and spin or
pseudospin $S$, for the $n=0$ and $n=1$ LLs at several fractions.  The
$n=0$ results are identical for GaAs and graphene.  The near perfect
overlaps in the last column of the table indicate that the FQHE in the
$n=1$ graphene LL also strongly resembles that in the $n=0$ (lowest)
graphene LL.  (The analogous overlaps are rather low for the FQHE
states in the $n=0$ and $n=1$ LLs in GaAs \cite{Ambrumenil88}.) The
1/3 state is fully polarized, whereas the 2/5 and 2/3 are (pseudo)spin
singlet.  At $3/5$ the spin of the ground state differs from that of
the lowest LL for $N=8$; the existence or the nature of FQHE at this
fraction remains unclear at the moment.

\noindent {\bf Pseudoskyrmions:}
In GaAs quantum wells, the excitations of the $\nu=1$ state for
exactly zero Zeeman splitting are not simple particle-hole excitations
but spin textures called skyrmions \cite{Lee90,Sondhi,Moon}, in which
half of the spins are reversed.  However, the skyrmion size rapidly
decreases with increasing Zeeman energy; experimentally, skyrmions
typically have 3 to 5 flipped spins \cite{Fertig94,Barrett95}.  No
skyrmions occur at $\nu=3, 5, \ldots$.  CF skyrmions are believed to
be relevant near $\nu=1/3$ at very small Zeeman energies
\cite{Kamilla96a,Wojs02,Leadley97}.

In contrast, for the state in which one of the two degenerate levels
of the $n=0$ graphene manifold is fully occupied (which produces zero
Hall conductance), the excitations ought to be large pseudoskyrmions.
In exact diagonalization studies, we find pseudoskyrmions also in the
$n=1,2$ graphene LLs, where the addition of one particle or hole to
the fully pseudospin polarized state produces a pseudospin singlet
state.  No such behavior is seen for $n\geq 3$; here the excitation is
fully pseudospin-polarized ($S=N/2$) on the quasihole side and has a
single pseudospin reversed ($S=N/2 - 1$) on the quasiparticle
side. Fig.~\ref{allgraphs}(b) depicts the $N$ dependence of the gap to
creating a pair of pseudoskyrmion and anti-pseudoskyrmion, computed by
exact diagonalization. (We follow the convention of Ref.\
\cite{Sondhi} to define the pseudoskyrmion gap in terms of ``neutral''
quasiparticle/quasihole energies \cite{GM}.)  Extrapolation to the
thermodynamic limit yields $\Delta_1^{(0)}=0.606(15)$,
$\Delta_1^{(1)}=0.126(7)$, and $\Delta_1^{(2)}=0.18(1)$.  The gap in
the $n=0$ LL is consistent with $\sqrt{\pi/8}\, (e^2/\epsilon
\ell_B)$, half the energy required to create an ordinary particle-hole
pair excitation \cite{Sondhi} .  We note that the pseudospin texture
in the $n=0$ graphene LL can be imaged directly by scanning tunneling
microscopy, since an electron's pseudospin determines on which
sublattice it resides.

Fig.~\ref{allgraphs}(c-d) shows the gaps as a function of the number
of particles for several fractional filling factors in the $n$th
graphene LL.  The lowest-energy charged excitations at $\nu^{(1)}=1/3$
are CF pseudoskyrmions (pseudo-spin singlet state), with an excitation
energy of 0.017, to be compared to 0.096 for a (pseudospin reversed)
particle-hole pair of composite fermions.  (The latter gap is greater
than in the $n=0$ LL because the $V_1$ pseudopotential is greater in
the $n=1$ than in the $n=0$ LL.)  The gaps at filling factors 2/5 and
2/3 involve pseudospin reversal for composite fermions.  For 2/5 the
gaps, $\Delta_{2/5}^{(0)}=0.051(1)$ and $\Delta_{2/5}^{(1)}=0.062(1)$,
were obtained from the trial wave functions of the CF theory,
evaluated by a procedure described in Ref.\ \cite{secondll}.

We summarize our principal conclusions.  The lowest LL FQHE of
graphene in the large Zeeman energy limit is equivalent to the lowest
LL FQHE in GaAs in the zero Zeeman energy limit, terminating into a
pseudospin-singlet Fermi sea at half filling.  The effective
interaction in Eq.~(\ref{effective}) is shown to be more favorable to
CF formation in the $|n|=1$ LL of graphene than in the $n=1$ LL of
GaAs.  The gaps at $\nu^{(1)}=1/3,\;2/5$ are calculated and FQHE is
predicted at $\nu^{(1)}=2/3$ due to reverse flux attachment.  In
contrast to GaAs, skyrmions are predicted to occur at $\nu^{(n)}=1,3$
in the $n=0,1,2$ LLs.

While completing the manuscript, we learned from Philip Kim of the
observation of new integral plateaus by Zhang {\it et
  al.}~\cite{Zhang}.  Several recent theory preprints have examined
similar topics, including a field theory approach that obtains results
similar to ours~\cite{Yang06} and derivations of the Coulomb
pseudopotentials for graphene~\cite{others}.  Support by the National
Science Foundation under grants no.\ DMR-0240458, DMR-0305035 and
ECS-0609243 is gratefully acknowledged.


\vspace{-3 mm}

\begin{thebibliography}{99}


\bibitem{Novoselov}
K.S.~Novoselov
{\em et al.}, Nature \textbf{438}, 197 (2005); Y.~Zhang {\em et al.}, Nature \textbf{438}, 201 (2005);


\bibitem{shon-ando} N.~H.~Shon and T. Ando, J. Phys. Soc. Jpn. {\bf 67}, 2421 (1998).

\bibitem{Semenoff} G.W.~ Semenoff, Phys. Rev. Lett. \textbf{53}, 2449 (1984).

\bibitem{Tsui82} D.C. Tsui, H.L. Stormer, and A.C. Gossard,
Phys. Rev. Lett. {\bf 48}, 1559 (1982).


\bibitem{Wu93} X.G. Wu, G. Dev, and J.K. Jain, Phys. Rev. Lett.  {\bf
71}, 153 (1993); K. Park and J.K. Jain, {\em ibid.} {\bf 80}, 4237 (1998).

\bibitem{Jain89}  J.K. Jain, Phys. Rev. Lett. {\bf 63}, 199 (1989).

\bibitem{notation} The index $n$ is used for both electron and CF Landau levels;
the meaning should be clear from the context.  Also, $\nu^{(n)}$ denotes the
filling factor in the $n$th LL, with the lower LLs assumed to be filled and inert.

\bibitem{Rezayi2} E.H. Rezayi, Phys. Rev. B {\bf 36},
5454 (1987); S.S. Mandal and J.K. Jain, {\em ibid.} {\bf 63},
201310 (2001).

\bibitem{Kalmeyer92} V. Kalmeyer and S.C. Zhang, Phys. Rev. B {\bf 46}, 
R9889 (1992); B.I. Halperin, P.A. Lee, and N. Read, Phys. Rev. B {\bf
47}, 7312 (1993).


\bibitem{Willett93} R.L. Willett {\em et al.},
Phys. Rev. Lett. {\bf 71}, 3846 (1993);  W.~Kang {\em et al.}, {\em ibid.} {\bf 71}, 3850 (1993); V.J. Goldman {\em et al.},  {\em ibid.}  {\bf 72}, 2065 (1994);
J. H. Smet {\em et al.},   {\em ibid.}  {\bf 77}, 2272 (1996).


\bibitem{Park} K. Park {\em et al.}, Phys. Rev. B {\bf 58}, R10167 (1998);
K. Park and J.K. Jain, Phys. Rev. Lett. {\bf 80}, 4237 (1998).


\bibitem{Ambrumenil88} N. d'Ambrumenil and A.M. Reynolds, J.
Phys. C: Solid State Phys. {\bf 21}, 119 (1988).

\bibitem{Lee90} D.-H. Lee and C. L. Kane
Phys. Rev. Lett. {\bf 64}, 1313 (1990); 

\bibitem{Sondhi} S.L. Sondhi {\em et al.}, Phys. Rev. B {\bf 47}, 16419 (1993).

\bibitem{Moon}
K. Moon {\em et al.}, Phys. Rev.  B {\bf 51}, 5138 (1995).

\bibitem{Fertig94} H.A. Fertig {\em et al.},  Phys. Rev. B {\bf 50}, R11018 (1994).

\bibitem{Barrett95} S. E. Barrett {\em et al.}, Phys. Rev. Lett. {\bf 74}, 5112 (1995); A. Schmeller {\em et al.}, {\em ibid.} {\bf 75}, 4290 (1995); E. H. Aifer, B. B. 
Goldberg, 
and D. A. Broido,  {\em ibid.} 
{\bf 76}, 680 (1996).

\bibitem{Kamilla96a} R.K. Kamilla, X.G. Wu, and J.K.
Jain, Phys. Rev. Lett.  {\bf 76}, 1332 (1996).

\bibitem{Wojs02} A. W\'{o}js and J. J. Quinn, Phys. Rev. B {\bf 66}, 
045323 (2002).

\bibitem{Leadley97} D. R. Leadley {\em et al.},
Phys. Rev. Lett. {\bf 79}, 4246 (1997).

\bibitem{GM} A.~H.~MacDonald and S.~M.~Girvin, Phys. Rev. B \textbf{34}, 5639 (1986).

\bibitem{secondll}
C.~T\H oke {\em et al.}, Phys. Rev. B \textbf{72}, 125315 (2005).

\bibitem{Zhang} Y.~Zhang {\em et al.}, Phys. Rev. Lett. \textbf{96}, 136806 (2006).

\bibitem{Yang06} K. Yang, S. Das Sarma, and A.H. MacDonald, cond-mat/0605666 (2006).

\bibitem{others}
K.~Nomura and A.~H.~MacDonald, Phys. Rev. Lett. {\bf 96}, 256602 (2006);
V.~.M.~Apalkov and T.~Chakraborty, cond-mat/0606037 (2006);
M.~O.~Goerbig, R.~Moessner, and B.~Dou\c{c}ot, cond-mat/0604554 (2006).


\end{thebibliography}
\end{document}